\documentclass[a4paper,fleqn,usenatbib]{mnras}

\usepackage{newtxtext,newtxmath}
\usepackage[T1]{fontenc}
\usepackage{ae,aecompl}

\usepackage{graphicx}	% Including figure files
\usepackage{amsmath}	% Advanced maths commands
\usepackage{amssymb}	% Extra maths symbols

\title[Hot Jupiters at High Resolution]{Hot Jupiter Atmospheric Flows at High Resolution}

\author[K. Menou]{
Kristen Menou,$^{1, 2, 3}$
\\
$^{1}$  Physics \& Astrophysics Group, Dept.   of  Physical  \&  Environmental  Sciences,  University  of  Toronto  Scarborough,\\   1265 Military Trail, Toronto, Ontario, M1C 1A4, Canada \\
$^{2}$ Dept.  of Astronomy \& Astrophysics, University of Toronto.
50 St.  George Street, Toronto, Ontario, M5S 3H4, Canada 
\\
$^{3}$ Dept.  of Physics, University of Toronto.
60 St George Street, Toronto, Ontario, M5S 1A7, Canada \\
}

\date{Accepted XXX. Received YYY; in original form ZZZ}

\pubyear{2019}

\begin{document}
\label{firstpage}
\pagerange{\pageref{firstpage}--\pageref{lastpage}}
\maketitle

\begin{abstract}
Global Circulation Models (GCMs) of atmospheric flows are now routinely used to interpret observational data on Hot Jupiters. Localized "equatorial $\beta$-plane" simulations by  \cite{2016A&A...591A.144F} have revealed that a barotropic (horizontal shear) instability of the equatorial jet appears at horizontal resolutions beyond those typically achieved in global models; this instability could limit wind speeds and lead to increased atmospheric variability. To address this possibility, we adapt the  computationally efficient, pseudo-spectral PlaSim GCM, originally designed for Earth studies, to model Hot Jupiter atmospheric flows and validate it on the \cite{2011MNRAS.413.2380H} reference benchmark. We then present high resolution global models of HD209458b, with horizontal resolutions of T85 (128x256) and T127 (192x384). The barotropic instability phenomenology found in $\beta$-plane simulations is not reproduced in these global models, despite comparably high resolutions. Nevertheless, high resolution models do exhibit additional flow variability on long timescales (of order 100 planet days or more), which is absent from the lower resolution models. It manifests as a breakdown of north-south symmetry of the equatorial wind. From post-processing the atmospheric flows at various resolutions (assuming a cloud-free situation), we show that the stronger flow variability achieved at high resolution does not translate into noticeably stronger dayside infrared flux variability. More generally, our results suggest that high horizontal resolutions are not required to capture the key features of hot Jupiter atmospheric flows.  
\end{abstract}


%%%%%%%%%%%%%%%%%%%%%%%%%%%%%%%%%%%%%%%%%%%%%%%%%%

%%%%%%%%%%%%%%%%% BODY OF PAPER %%%%%%%%%%%%%%%%%%

\section{Introduction}
Hot Jupiters atmospheres have been extensively characterized by observational campaigns and, as a result, they provide some of the best laboratories available to us for studying the physics of exoplanet atmospheres \cite{2007arXiv0706.1047C, 2010ARA&A..48..631S,2010RPPh...73a6901B, 2016SSRv..205..285M, 2018haex.bookE.116P}. The atmospheric flows that develop on these planets, with permanent day and night sides, are the subject of ongoing modelling efforts. While we know and have an understanding of what drives winds in these atmospheres \citep{2011ApJ...738...71S, 2018ApJ...869...65H},  the dominant form of dissipation for wind kinetic energy is still the subject of intense debate. Indeed new physics is at play in these atmospheres: shocks \citep{2010ApJ...725.1146L, 2013MNRAS.435.3159D, 2016A&A...591A.144F} MHD effects \citep{2010ApJ...714L.238B, 2010ApJ...719.1421P, 2012ApJ...745..138M, 2018AJ....155..214T} and vertical transport \citep{2019MNRAS.485L..98M} could all play a role in limiting the wind speeds on these planets.

In their study of shocks in hot Jupiter atmospheres, \cite{2016A&A...591A.144F} found that a barotropic horizontal shear instability develops in a deep model of the specific hot Jupiter HD209458b. The phenomenology of this instability, and the time variable flow that results from it, is reminiscent of the barotropic instability discussed by \cite{2009ApJ...700..887M} in their shallow hot Jupiter model \citep[see also][]{2011MNRAS.413.2380H}. To this day, however, a similar shear instability has not manifested in any of the deep global hot Jupiter flow models published in the literature. \cite{2016A&A...591A.144F} results suggest that this could be due to insufficient resolution in typical deep global model since the instability does not occur in their simulations until a demanding latitudinal resolution threshold is met (see their Figure 4). The existence of such an instability is important in (i) offering a novel avenue to limit wind speeds in hot Jupiter atmospheres and (ii) in the possibility that it could manifest observationally in the form of atmospheric photometric and/or spectroscopic variability. Recently, \cite{2019arXiv191009523K} have reviewed the observational evidence on hot Jupiter variability and presented a detailed exploration of the level of variability expected from global circulation models.

Here we reconsider the horizontal shear instability problem with PlaSim-Gen,  an adaptation of the fast PlaSim Earth system simulator to the case of deep atmospheres such as those of hot Jupiters. Our work complements the recent study by \cite{2019arXiv191009523K} . The plan of this article is as follows. In \S2, we describe our validation of the PlaSim-Gen model. In \S3, we present high horizontal resolution models of HD209458b,  which allow us to address the barotropic shear phenomenology of \cite{2016A&A...591A.144F}. We also present explicit post-processed day-side variability diagnostics of our atmospheric flow models. We find no corroborating evidence for the \cite{2016A&A...591A.144F} phenomenology and establish that the flow variability exhibited in our high-resolution models does not translate into any significant additional observational variability in thermal emission. We conclude in \S4.

\section{PlaSim-Gen Model Validation}

As described in further details in Appendix A,  PlaSim-Gen  is an adaptation of PlaSim to model deep atmospheres such as those of hot Jupiters.  In PlaSim-Gen, the atmosphere is decoupled from the surface, it is assumed dry and various Earth specific modules have been turned off. In the present version, the radiative forcing has been simplified by implementing Newtonian relaxation to precomputed radiative equilibrium profiles (see Appendix A for details).

We validate our implementation by reproducing the deep HD209458b benchmark of \cite{2011MNRAS.413.2380H}. Our radiative relaxation profiles follow the polynomial fits of \cite{2011MNRAS.413.2380H}. We perform the validation at a moderate resolution of T31L30, which is commensurate with other spectral-core models for hot Jupiters published in the literature \citep[e.g.][]{2011MNRAS.413.2380H, 2012ApJ...750...96R}. 
A model time step MPSTEP = 180s and dissipation timescale TDISS = $5 \times 10^{-3}$ were adopted. We run models up to 1200 planet days, which is longer than the typical runtime of models in the literature. Other model parameters are listed in Table 1 (see also Table A1 in Appendix A).

The top panels in Figure 1 and Figure 2 show, respectively, the zonal mean zonal wind profile and a representative horizontal temperature slice at the 260 mb level for our T31L30 validation model.  We find that the outcomes from our validation model compare well to the reference results presented by \cite{2011MNRAS.413.2380H}.

\section{PlaSim-Gen High Resolution Models}

Next, we move to high resolution versions of this HD209458b model. We focus on resolutions T85L30 and T127L30, which means that we increase the horizontal resolution while keeping the vertical resolution unchanged. This choice is dictated by our attempt to compare directly to the increased latitudinal resolution results discussed by \cite{2016A&A...591A.144F}.  We run our models with MPSTEP = 60s (T85) and MPSTEP = 45s (T127), and values of the hyperdissipation timescale  applied to all dynamical variables TDISS = $5 \times 10^{-4}$ (T85) and TDISS = $1 \times 10^{-4}$ (T127). We find that PlaSim-Gen is fairly efficiently parallelized,  up to 32 threads, allowing us to run a T127 model for 1200 planet days in less than a week of wall time on a modern workstation.

\begin{figure}
	%\includegraphics[width=\columnwidth]{UZ_profile_noVD.pdf}
	\includegraphics[width=\columnwidth]{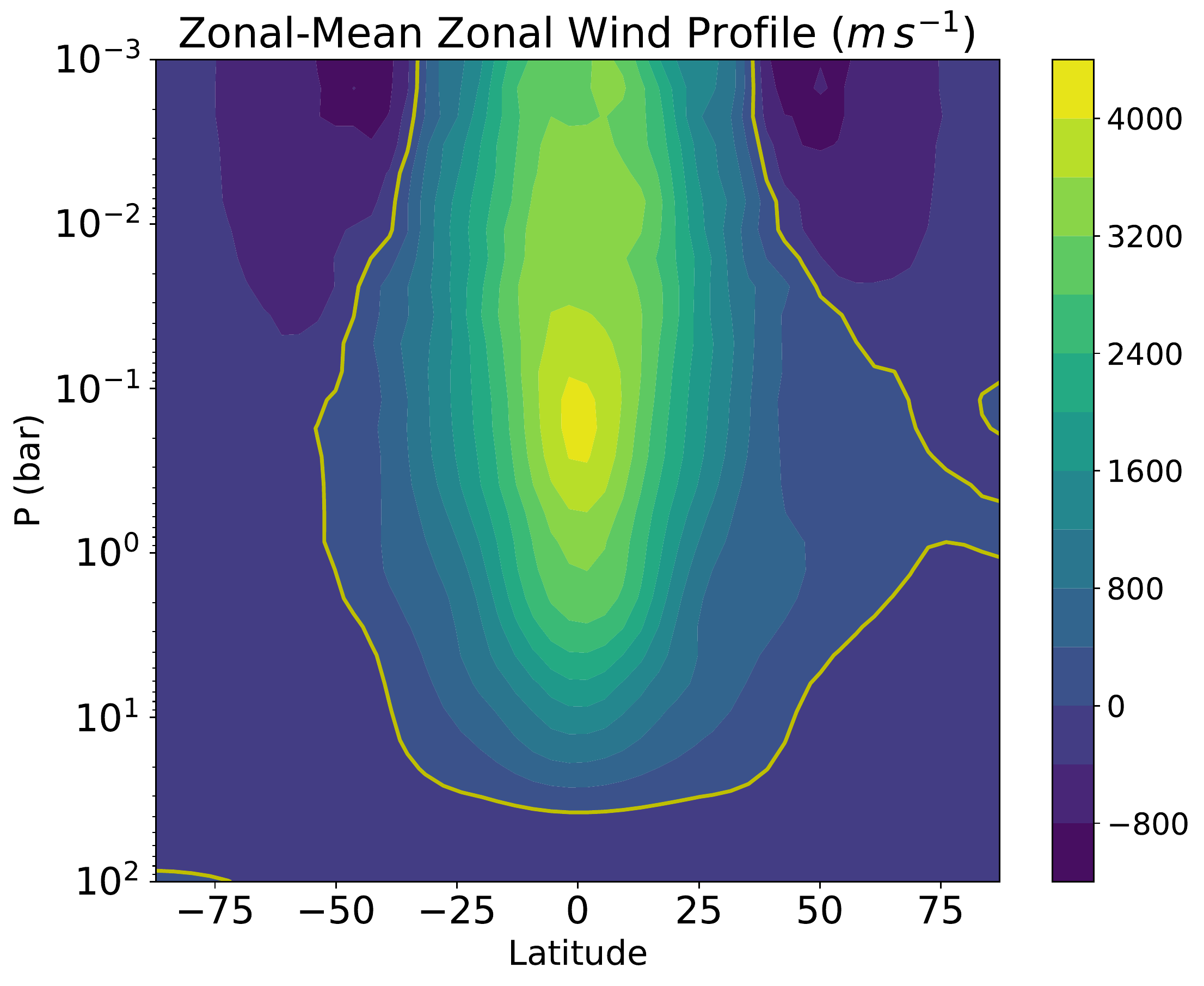}
	\includegraphics[width=\columnwidth]{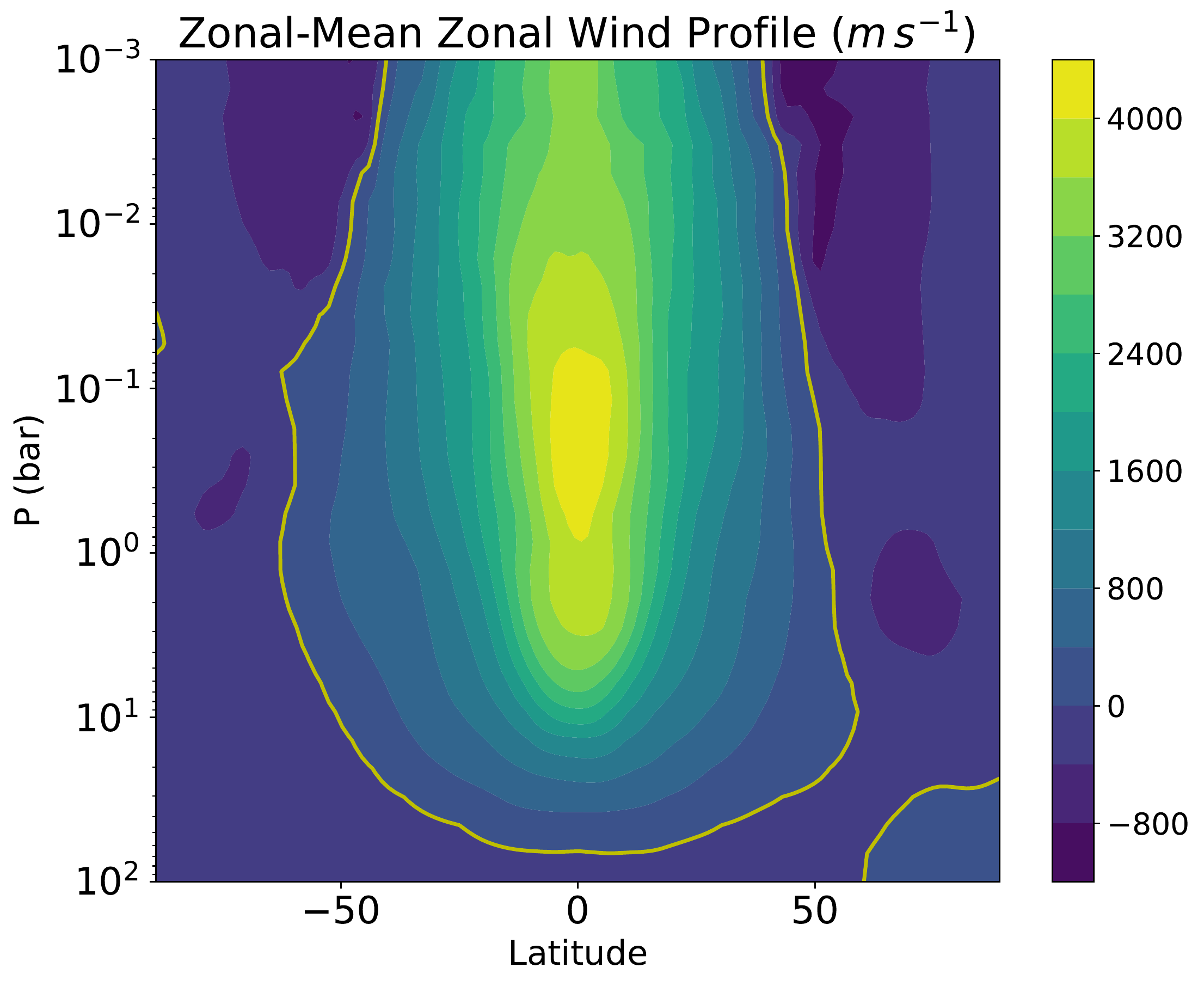}
	\includegraphics[width=\columnwidth]{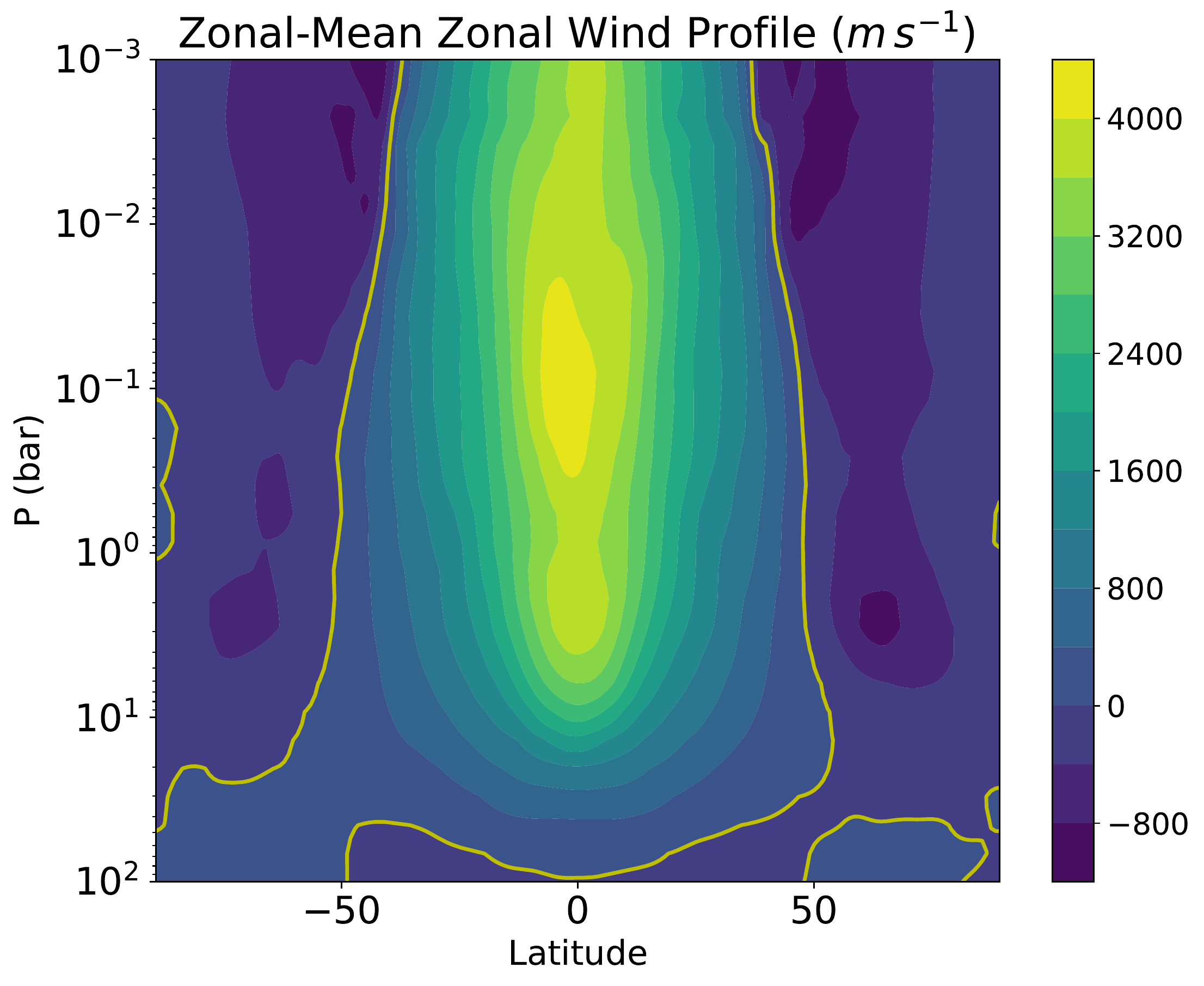}
    \caption{Zonally-averaged zonal wind profile at planet day 1200 in our T31L30 (top), T85L30 (middle) and T127L30 (bottom) models of HD209458b. Compare to Figure 12 in Heng et al. (2011).}
    \label{fig:one}
\end{figure}

\begin{figure}
	%\includegraphics[width=\columnwidth]{UZ_profile_noVD.pdf}
	\includegraphics[width=\columnwidth]{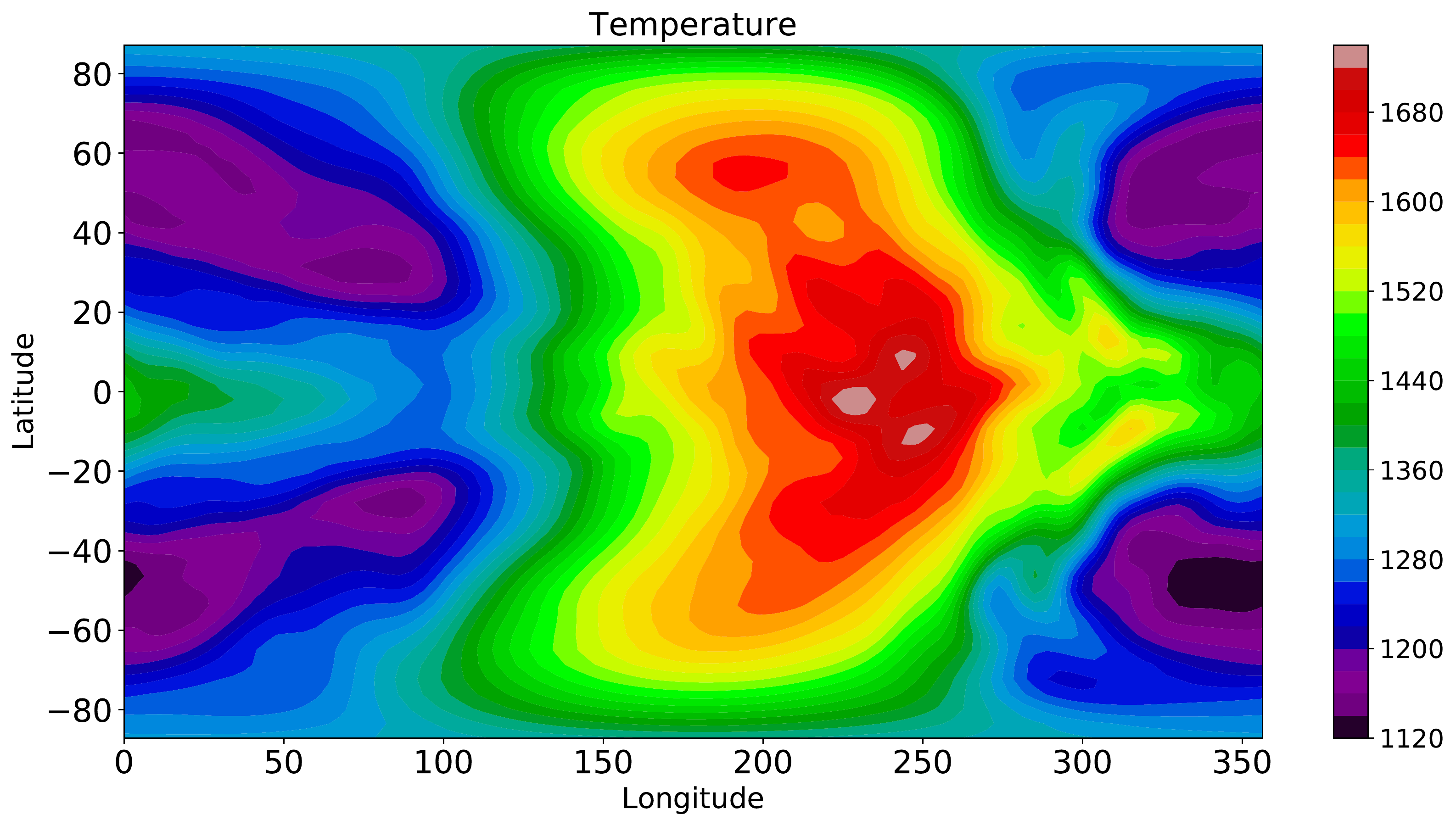}
	\includegraphics[width=\columnwidth]{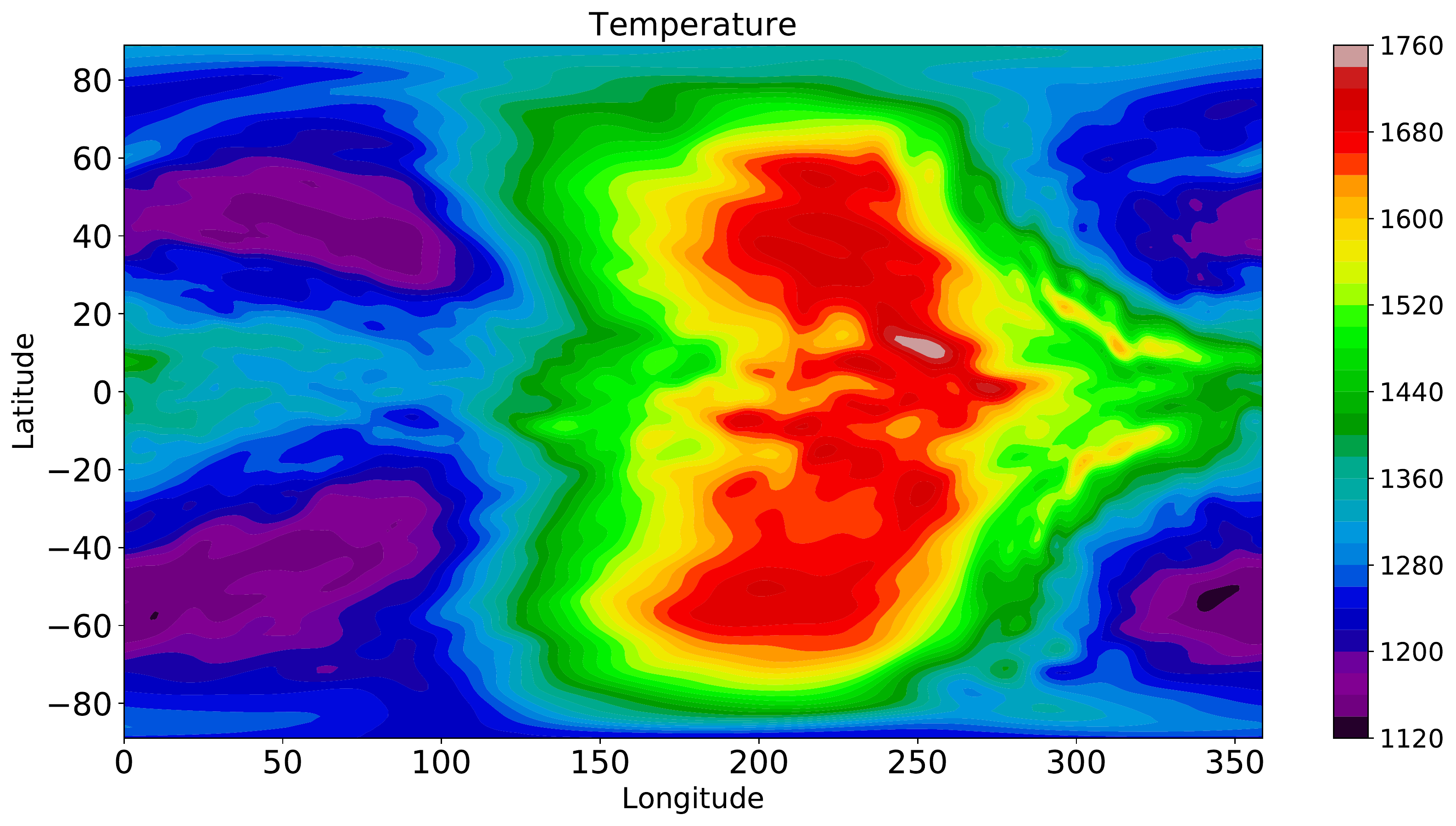}
	\includegraphics[width=\columnwidth]{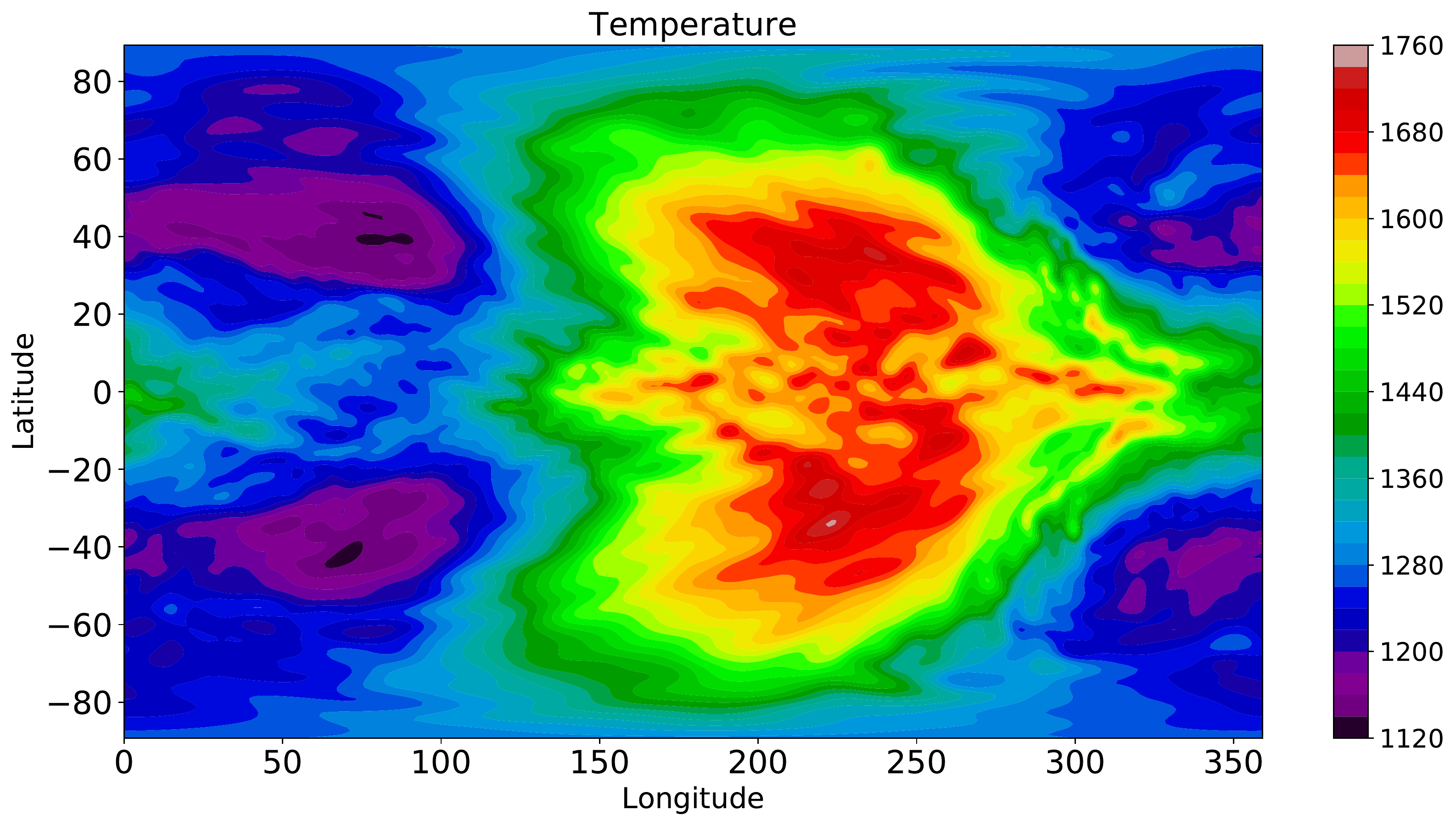}
    \caption{Temperature maps at the 260mb pressure level at planet day 1200 in our T31L30  (top), T85L30 (middle) and T127L30 (bottom) models of HD209458b. Compare to Figure 9 in Heng et al. (2011)}
    \label{fig:two}
\end{figure}

The middle panel in figures 1 and 2 show our T85L30 results while the lower panels  show the T127L30 results. By and large, higher-resolution models reproduce the results of lower resolution versions (T31L30), with smaller-scale flow features that do not seem to impact much the global flow properties. This has already been noticed by \cite{2011MNRAS.413.2380H} and \cite{2013ApJ...770...42L}.

While the zonal wind structure, wind speeds and temperature maps are largely consistent across resolutions, careful examination reveals a deeper penetration depth of the equatorial jet  and a somewhat larger north-south asymmetry (in both zonal wind and temperature maps), which seems absent from the lower T31L30 resolution runs.

These results are worth discussing in the context of the study of \cite{2016A&A...591A.144F}, who find a transition to a markedly different regime of circulation at latitudinal resolutions in excess of 64-128 cell elements (see their Figure 4). Our T85 and T127 runs do reach the latitudinal regime at which one might expect the dynamical transition witnessed by \cite{2016A&A...591A.144F} to manifest. However, as exemplified in the lower two panels of Figure 2, even at high resolution, we do not find any evidence for the oscillatory/cyclic behavior of the equatorial regions reported by \cite{2016A&A...591A.144F}.

It is conceivable that our results on equatorial flow stability and lack of variability 
are  sensitive to the level of hyperdissipation adopted in our models, as has already been discussed by \cite{2011MNRAS.413.2380H} for example. To address this possibility, we have run additional T31 and T85 models with different levels of hyperdissipation.  We label T31\_v1 and T85\_v1 our previously described models with TDISS $=5 \times 10^{-3}$ and $=5 \times 10^{-4}$, respectively. We also ran models with higher levels of hyperdissipation: T31\_v2 with TDISS $=1 \times 10^{-3}$ and T85\_v2 with TDISS $=1 \times 10^{-4}$ (i.e.,  each with five times faster hyperdissipation than in v1 models). Overall we find little differences in atmospheric flow variability between or v1 and v2 models.

To better quantify our various claims, we inspect the time viability of the equatorial wind speed at the 50mb  level, which is one of the variability diagnostics studied by \cite{2016A&A...591A.144F}. Figure 3 shows planet-daily time-series of the equatorial zonal wind speed from planet day 100 to 1200 (avoiding transients from atmospheric spin-up in the first 100 planet days). There is little difference in the nature of the equatorial wind variability between the models with more or less hyper dissipation (at least at T31 or T85 resolutions).

There is however a noticeable difference in the nature of the variability in the high-resolution models (T85 and T127),  relative to the moderate resolution models (T31).  In particular, past 400 to 600 planet days,  there are larger variations, $\sim$ 10 to 20\%  in fractional amplitude, on timescales  $\sim$ 100 to 200 planet days, which are absent from the low resolution runs. 
While the emergence of this extra variability is intriguing in that it matches expectations in terms of the resolution requirements discussed by \cite{2016A&A...591A.144F}, we find that the phenomenology of this variability differs from that described by \cite{2016A&A...591A.144F}.

The significant shift in equatorial wind speeds,  at the $\sim$10 to 20\% level, occur
on much longer timescales in our high-resolution runs than the ones reported by \cite{2016A&A...591A.144F} ($\sim$ 
days).  There is also no clear indication of wiggliness of the equatorial wind in our models (see Fig. 2). Furthermore, careful examination of our models reveals
that the shifts in equatorial wind speed occur as a result of a minor, but noticeable, north-south
breaking of symmetry in our high-resolution runs, which positions the peak wind speed somewhat above or below the equator. Some evidence for this behaviour is revealed in the lower two panels of Figures 1 and 2, although the effect is small.

One also notices in Figure 1 that the equatorial zonal flow
reaches deeper in our T85, and even more so T127, models than in the T31 model. One possible interpretation for the overall phenomenology exhibited in our high-resolution models is that the atmospheric flow,  driven from the top, eventually makes contact with the model lower boundary. This would happen at late times,  past $\sim 500$ planet days, and more easily so in high resolution models, because these models better resolve the meridional circulation associated with the top-forced zonal flow and its gradual penetration with depth \citep[see, e.g.][]{2006Icar..182..513S}. We have not pursued this interpretation further since we believe that additional physical ingredients are needed first to adequately model the deep layers at pressures 10-100 bars and their interaction with the deeper planetary interior \citep[see, e.g., the bottom drag included by][]{2013ApJ...770...42L, 2019arXiv191009523K}.

\section{Atmospheric Variability}

Variability in the atmospheric flow does not necessarily translate into observable photometric and/or spectroscopic variability. For example, small-scale variability may average out once the emission properties over an entire planetary hemisphere are considered.

\begin{figure}
	%\includegraphics[width=\columnwidth]{UZ_profile_noVD.pdf}
	\includegraphics[width=\columnwidth]{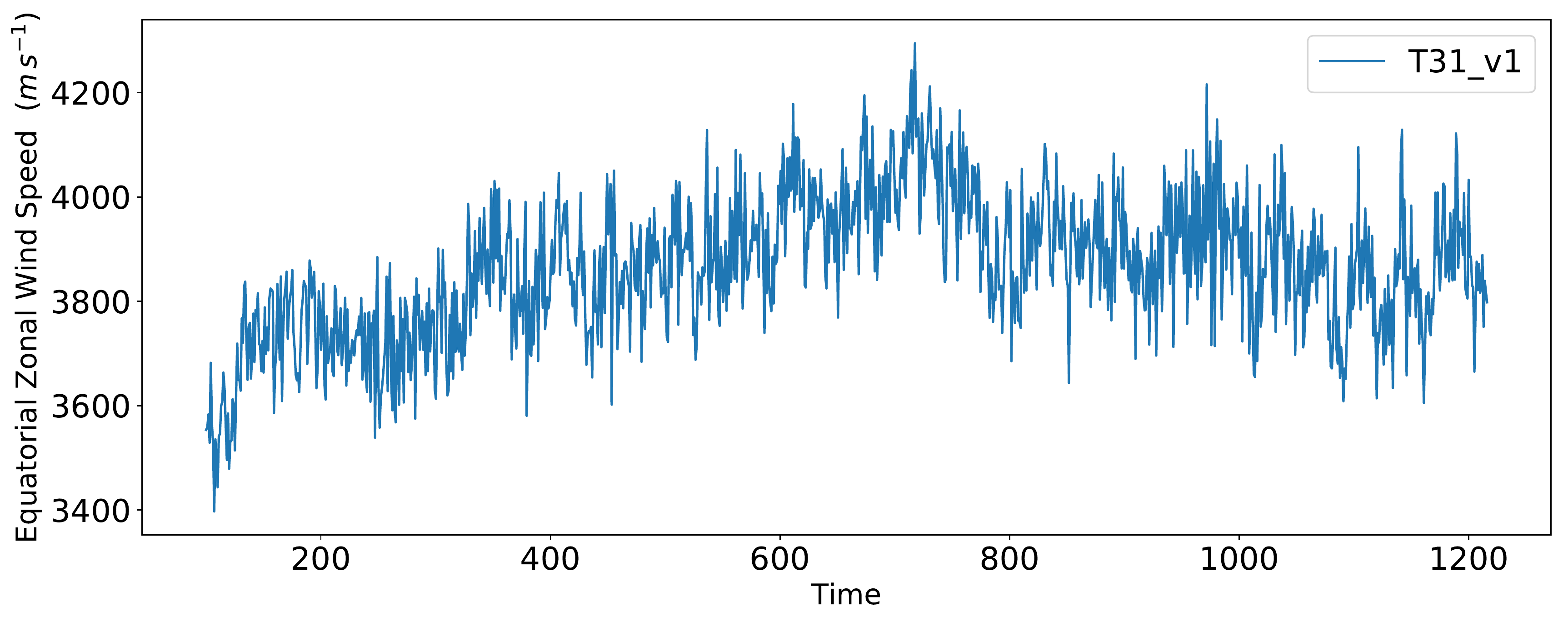}
	\includegraphics[width=\columnwidth]{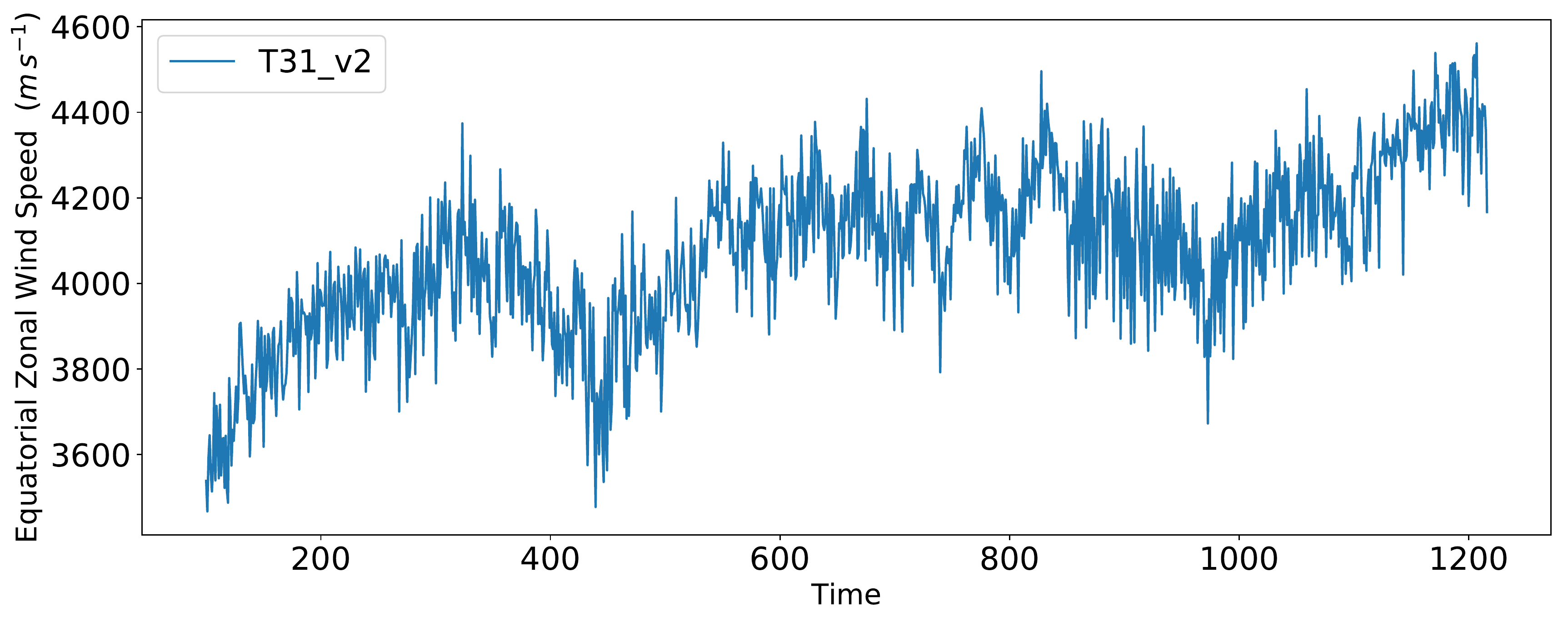}
	\includegraphics[width=\columnwidth]{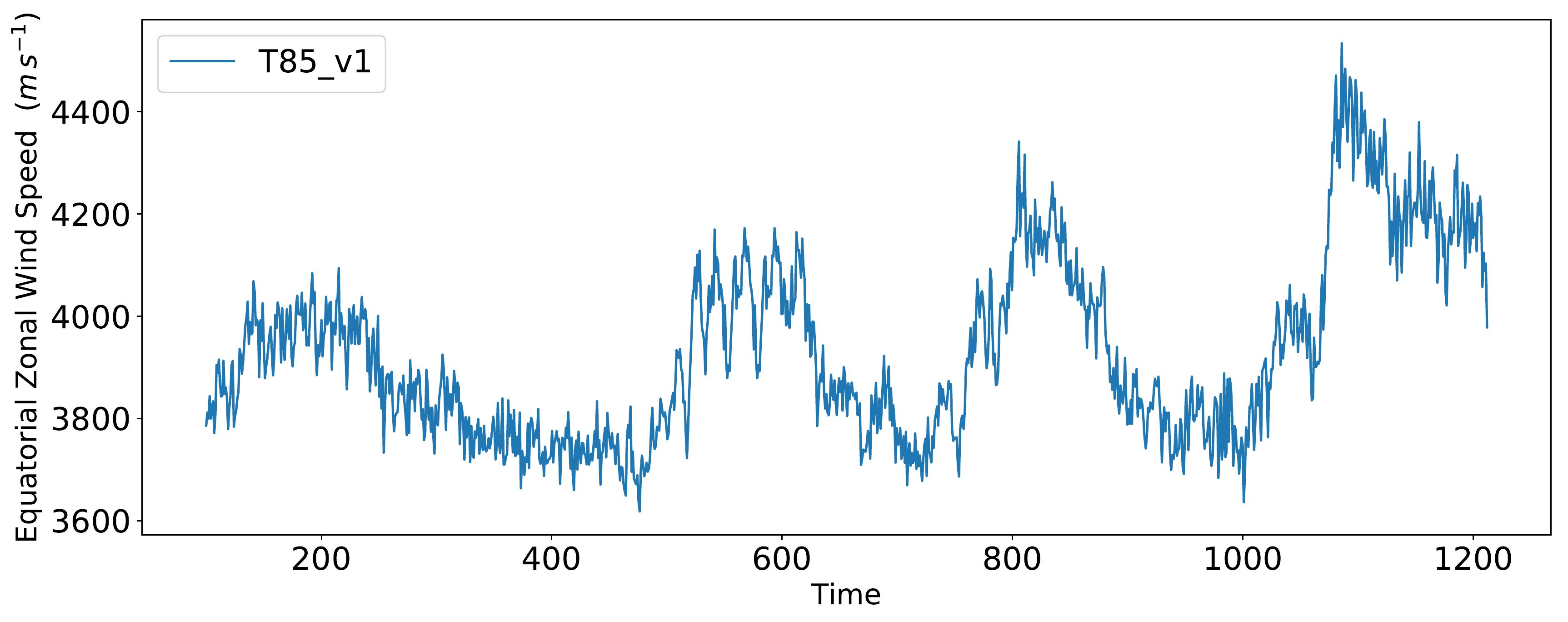}
	\includegraphics[width=\columnwidth]{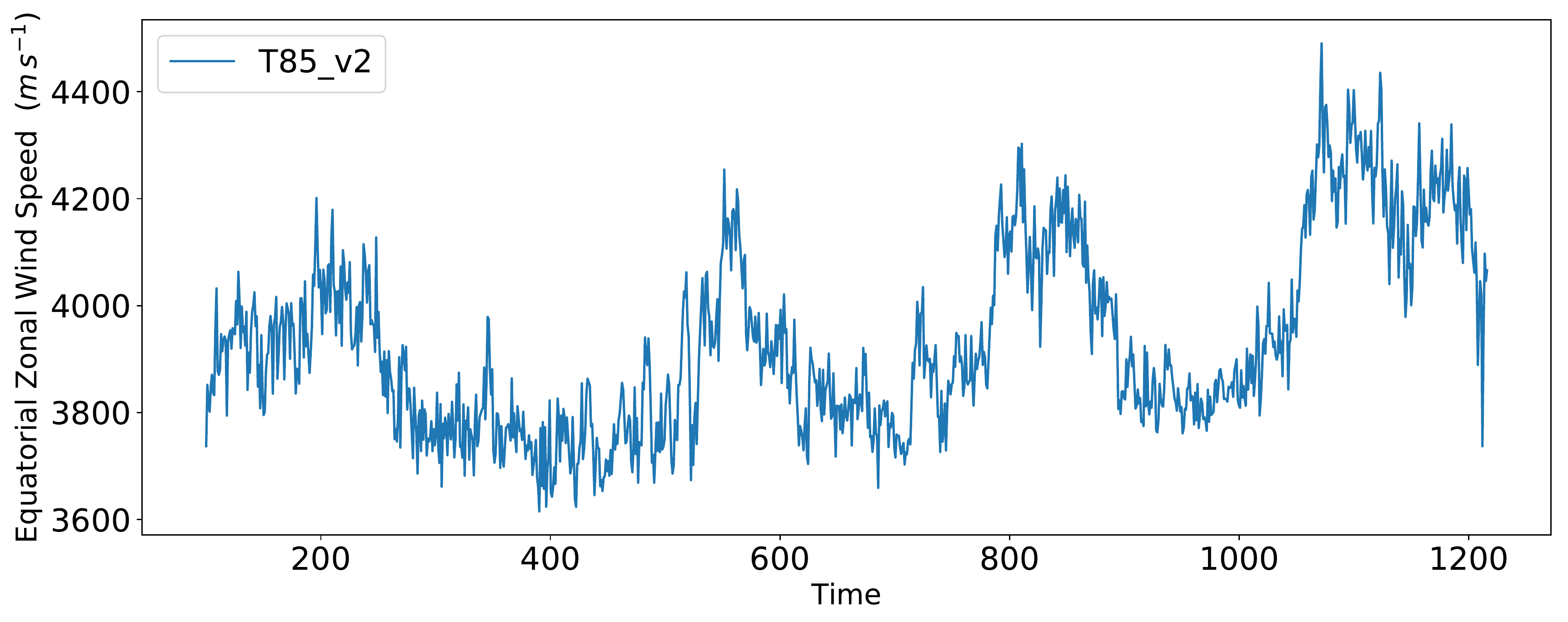}
	\includegraphics[width=\columnwidth]{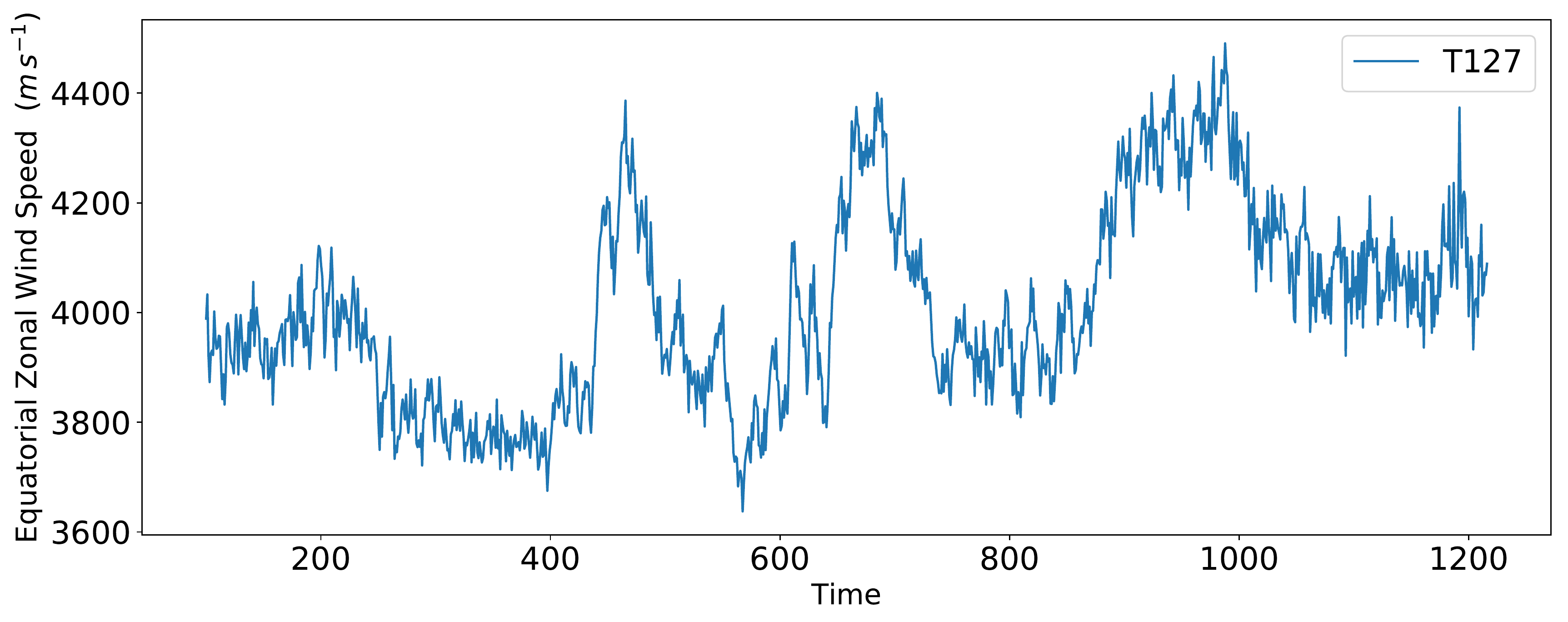}
    \caption{Equatorial zonal wind speed at the 53mb pressure level over planet days 100-1200. Compare to Figure 6 in Fromang et al. (2006).}
    \label{fig:three}
\end{figure}

To quantify the degree of photometric variability one might expect from the equatorial flow variability shown in Figure 3,  we have post-processed our models T31\_v1,  T85\_v1 and T127 with the petitRADTRANS  open source tool \cite{2019A&A...627A..67}. As a diagnostic, we compute dayside emission in two separate narrow bands. We individually process each model column, weighting its emission by the cosine of the angle away from the substellar point,  and then integrate all column contributions over the dayside hemisphere.

For concreteness, we adopt the same atmospheric composition parameters as the petitRADTRANS  default code example (0.74 H$_2$, 0.24 He, $10^{-3}$ H$_2$O, $10^{-2}$ CO, $10^{-5}$ CO$_2$, $10^{-6}$ CH$_4$, $10^{-5}$ Na and $10^{-6}$ K,  by mass). All other model parameters are chosen to match those our HD209458b model. The atmosphere is assumed to be cloud-free. To reduce the computational cost of processing from 16000 to 36000 dayside columns (for T85 and T127 resolutions, respectively), we focus on the emission in two representative narrow bands:  2.5-2.6 and 3.8-3.9 microns.

\begin{figure}
	%\includegraphics[width=\columnwidth]{UZ_profile_noVD.pdf}
	\includegraphics[width=\columnwidth]{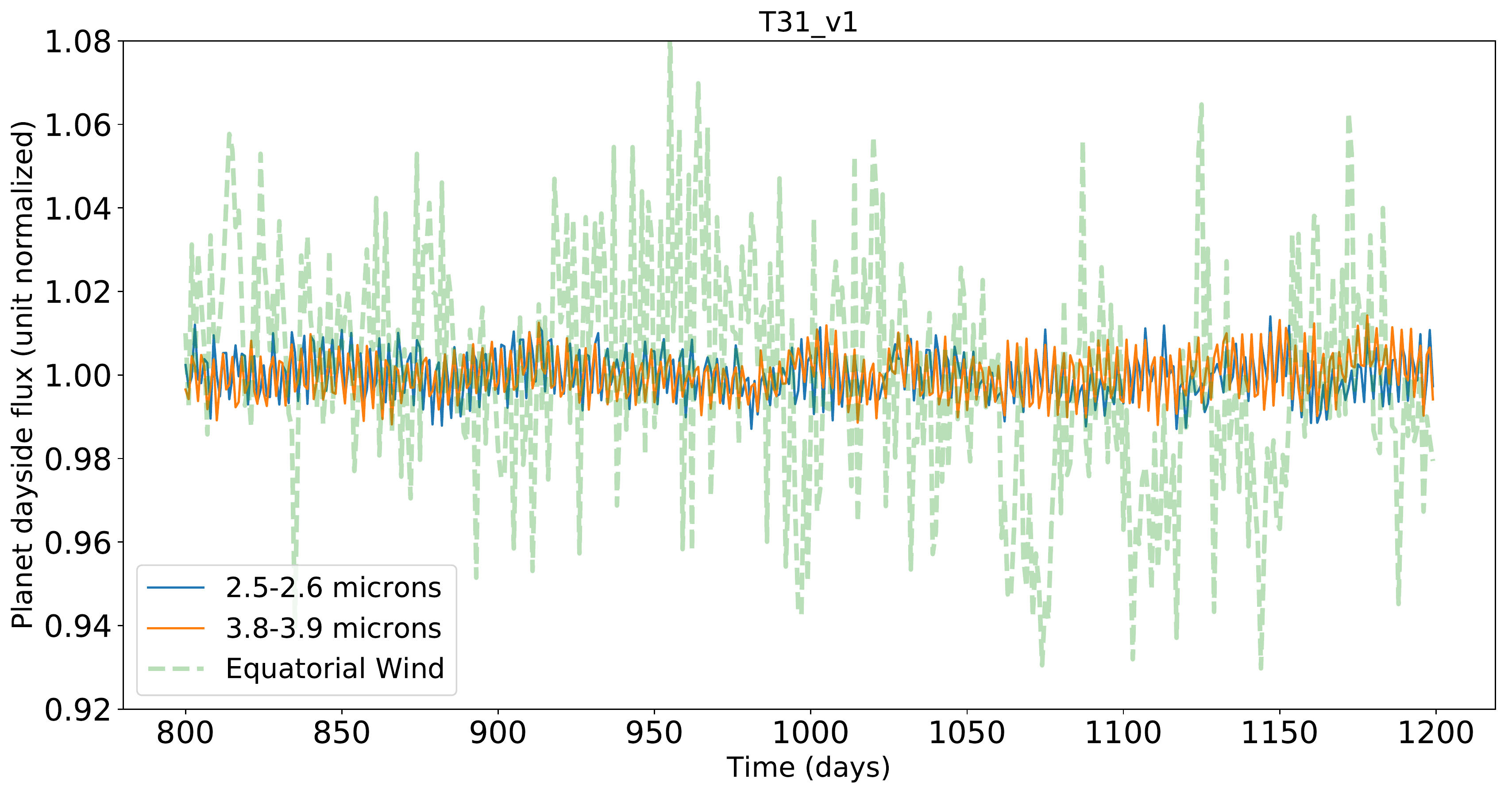}
	\includegraphics[width=\columnwidth]{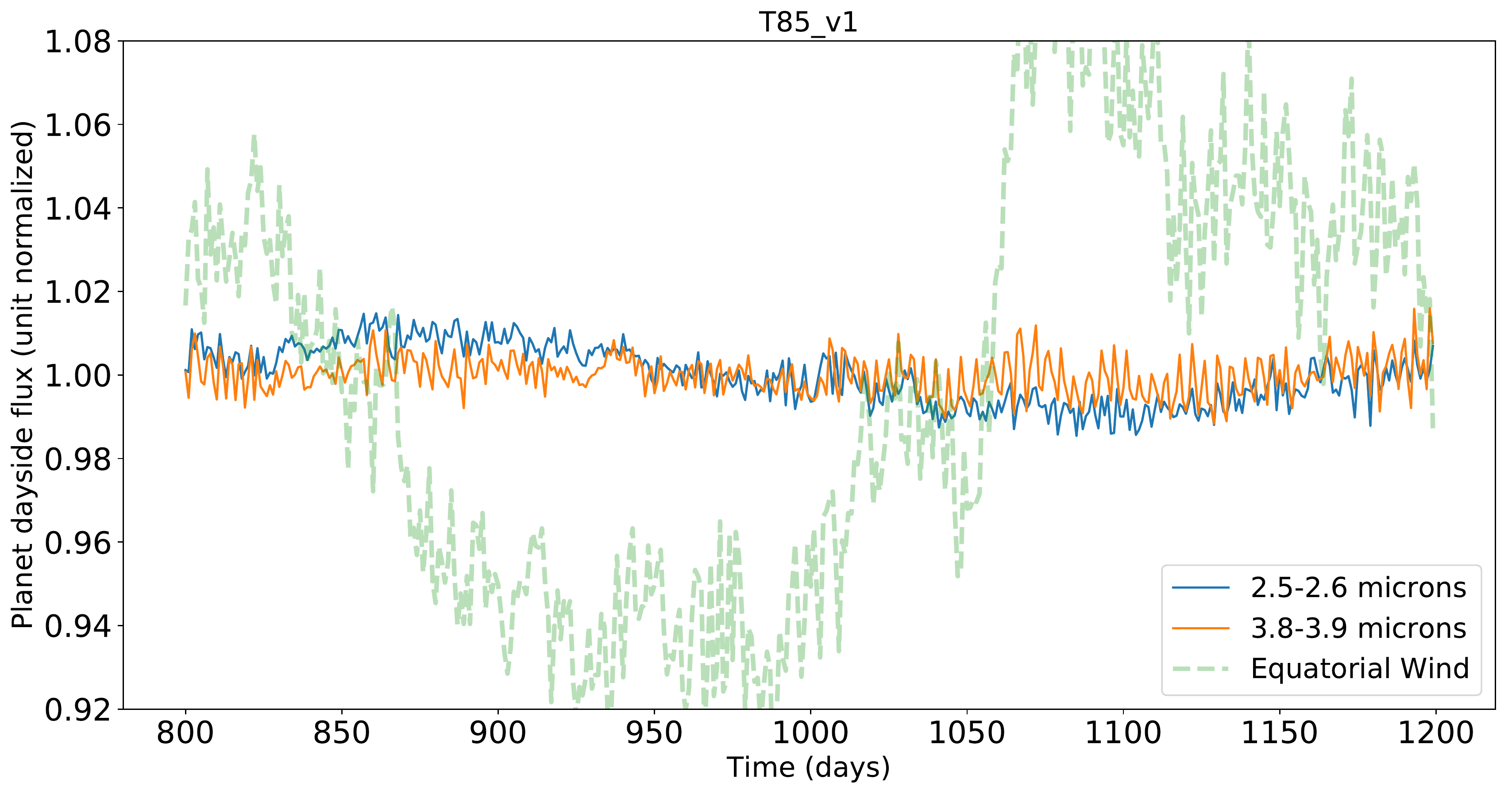}
	\includegraphics[width=\columnwidth]{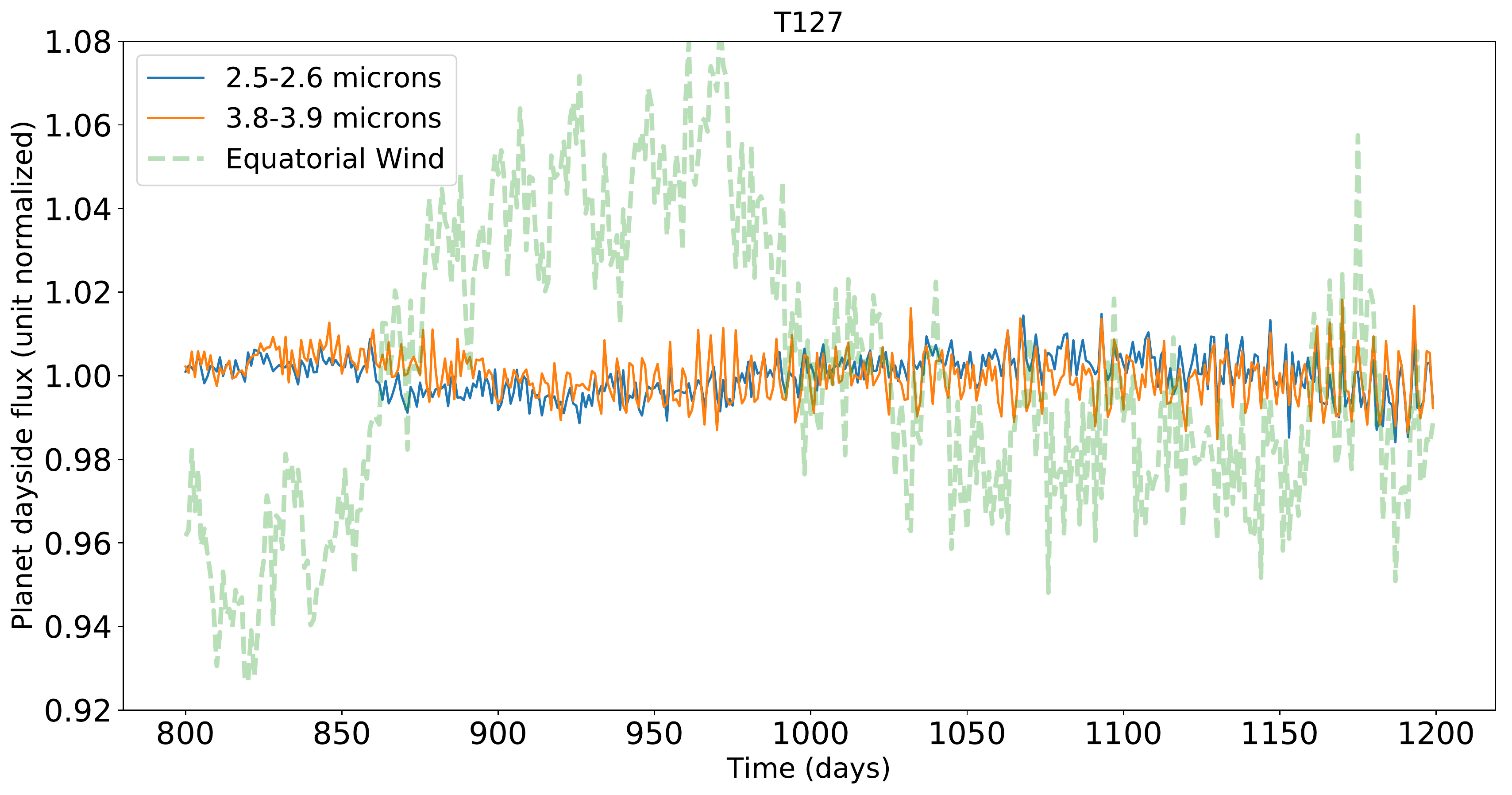}
    \caption{Variability over planet days 800-1200.}
    \label{fig:three}
\end{figure}

Figure 4 shows the dayside-integrated narrow-band photometric viability in these two bands (blue and orange solid lines),  with the equatorial wind from Figure 3  super-imposed as a green dash line. All quantities are shown in the time interval 800 to 1200 planet days and they have been normalized to emphasize relative variations. Comparing the top panel (T31\_v1) 
to the lower two panels (T85\_v1 and T127), it is clear that the larger equatorial wind variability
(at the 10 to 20\% level)  in the high-resolution models does not translate into any significant additional variability in the two narrow infrared bands of interest, relative to the modest resolution model (top panel).

The  dayside flux variations remain within the 2\% variability level for all three models.
This is consistent with our earlier observation that the larger equatorial wind variability at late times in the high-resolution models (Fig. 3) is largely caused by limited north-south shifts in the off-equatorial latitudinal location of the equatorial wind, which have limited impact on the dayside-averaged emission pattern of these atmospheres.

While not disproving the possibility of greater variability, e.g. when clouds are accounted for,
our results illustrate well the point that atmospheric flow variability does not necessarily translate into a more readily observable photometric/spectroscopic variability.

\begin{table}
	\centering
	\caption{Planet-Specific Model Parameters}
	\label{tab:example_table}
	\begin{tabular}{ccc} % four columns, alignment for each
		\hline
		Parameter & Value & Brief Description\\
		\hline
		GA & 8 & Gravitational acceleration ($m/s^2$)\\ 
		SIDEREAL\_DAY & $285120.0$ & 3.5 Earth day spin (s)\\ 
		ECCEN & 0 & Planet eccentricity\\ 
		OBLIQ & 0 & Planet obliquity\\ 
		RADIUS & $10^8$ & Planet Radius (m) \\ 
		PSURF & $10^7$ & Surface pressure (Pa)\\ 
		OOM & 5 & Pressure range: log10(PSURF/Ptop) \\ 
		AKAP & $0.286$ & Ratio of gas constant to specific heat\\ 
		GASCON & $3523$ & Atmospheric gas constant\\ 
		TGR & $2700$ & 'Ground' temperature (K)\\
		SOLAR\_DAY & $285120.0$ &  Solar day (s, unused)\\ 
		GSOL0 & $1.06 \times 10^6$ & Insolation flux ($W/m^2$, unused) \\ 
		\hline
	\end{tabular}
    \label{table:one}
\end{table}

\section{Discussion and Conclusions}

Superrotating equatorial jets like those realized in most simulations of hot Jupiter atmospheric flows are potentially subject to a barotropic (horizontal shear) instability \citep{1949JAtS....6..105K, 2006aofd.book.....V}.
While the barotropic instability was triggered in the shallow hot Jupiter model of \cite{2009ApJ...700..887M}, with $P_{\rm bottom}=1$ bar, none of the many deep hot Jupiter models ($P_{\rm bottom} \geq 100$ bar) published to date have shown the barotropic instability phenomenology. In this context, the findings by \cite{2016A&A...591A.144F}  that a barotropically  unstable  hot Jupiter flow can indeed be achieved in their deep model ($P_{\rm  bottom}=220$ bar), provided that the latitudinal resolution is high enough, are intriguing. Indeed, the reason why the barotropic instability did not manifest itself in other deep hot Jupiter models could simply be that these models did not have sufficient latitudinal resolution to resolve the instability onset. 

To address this possibility explicitly, we have built models at horizontal resolutions well beyond those typically achieved in global models of hot Jupiter atmospheric flows. We have found that the barotropic instability does not manifest in our deep global models, even at high horizontal resolution. While the atmospheric flow exhibits increased variability at high resolution, and at late times, we associate this feature to the deeper penetration of the equatorial jet observed at higher resolution (see Figure 1) and not the barotropic instability. 

Our findings raise a number of questions. First,  while \cite{2017A&A...604A..79M} and \cite{2019arXiv191010760M}  have argued that very long integration times are needed to reach a balanced state in deep model layers for hot Jupiter conditions, our results suggest that the exact requirement may be resolution-dependent (see Figure 1) and perhaps model-dependent as well (e.g.spectral versus grid-based algorithm).

Second, the evidence so far suggests that shallow enough global models can be susceptible to the barotropic instability, while deep enough models are not (or at least less so, within current computational limitations). This could simply be the result of the greater challenge there is in maintaining a barotropic (vertically-aligned) flow across a larger number of pressure scale heights. In this respect, we note that the low-latitude flow is clearly barotropic (vertically aligned) over a greater vertical extent in our T127 model (lower panel of Figure 1), relative to the other two panels, but this is not enough to trigger a horizontal shear instability.

Third, the reason for the emergence of the barotropic instability in the deep models of \cite{2016A&A...591A.144F}  remains unclear at this point. It could be related to the $\beta -$plane approximation (which is only valid in the close proximity of the equator) or perhap the compressible nature of the equations solved by \cite{2016A&A...591A.144F}, since compressible modes are filtered out from the hydrostatic equations solved in many published hot Jupiter models.  

To conclude, it would seem that the use of only moderate horizontal resolutions in global hot Jupiter models is supported by our results, in the sense that high resolutions are not strictly necessary to capture the key features of the atmospheric flow. On the other hand, our results also highlight how a suitable approach to model the deep atmospheric layers and their coupling to the planetary interior is still missing and in need of physical clarification.

\section*{Acknowledgements}
The author is grateful to Tad Komacek for comments on an early version of this manuscript. Computing resources were provided by the Canadian Institute for Theoretical Astrophysics at the University of Toronto. KM is supported by the National Science and Engineering Research
Council of Canada. This work has made extensive use of the following software packages: {\tt matplotlib, petitRADTRANS}.

We would furthermore like to acknowledge that our work was performed on land traditionally inhabited by the Wendat, the Anishnaabeg, Haudenosaunee, Metis, and the Mississaugas of the New Credit First Nation.

%%%%%%%%%%%%%%%%%%%%%%%%%%%%%%%%%%%%%%%%%%%%%%%%%%



\bibliographystyle{mnras}
\bibliography{hires_variability} % if your bibtex file is called example.bib

\begin{thebibliography}{}
\makeatletter
\relax
\def\mn@urlcharsother{\let\do\@makeother \do\$\do\&\do\#\do\^\do\_\do\%\do\~}
\def\mn@doi{\begingroup\mn@urlcharsother \@ifnextchar [ {\mn@doi@}
  {\mn@doi@[]}}
\def\mn@doi@[#1]#2{\def\@tempa{#1}\ifx\@tempa\@empty \href
  {http://dx.doi.org/#2} {doi:#2}\else \href {http://dx.doi.org/#2} {#1}\fi
  \endgroup}
\def\mn@eprint#1#2{\mn@eprint@#1:#2::\@nil}
\def\mn@eprint@arXiv#1{\href {http://arxiv.org/abs/#1} {{\tt arXiv:#1}}}
\def\mn@eprint@dblp#1{\href {http://dblp.uni-trier.de/rec/bibtex/#1.xml}
  {dblp:#1}}
\def\mn@eprint@#1:#2:#3:#4\@nil{\def\@tempa {#1}\def\@tempb {#2}\def\@tempc
  {#3}\ifx \@tempc \@empty \let \@tempc \@tempb \let \@tempb \@tempa \fi \ifx
  \@tempb \@empty \def\@tempb {arXiv}\fi \@ifundefined
  {mn@eprint@\@tempb}{\@tempb:\@tempc}{\expandafter \expandafter \csname
  mn@eprint@\@tempb\endcsname \expandafter{\@tempc}}}

\bibitem[\protect\citeauthoryear{{Baraffe}, {Chabrier}  \& {Barman}}{{Baraffe}
  et~al.}{2010}]{2010RPPh...73a6901B}
{Baraffe} I.,  {Chabrier} G.,   {Barman} T.,  2010, \mn@doi [Reports on
  Progress in Physics] {10.1088/0034-4885/73/1/016901}, \href
  {http://adsabs.harvard.edu/abs/2010RPPh...73a6901B} {73, 016901}

\bibitem[\protect\citeauthoryear{{Batygin} \& {Stevenson}}{{Batygin} \&
  {Stevenson}}{2010}]{2010ApJ...714L.238B}
{Batygin} K.,  {Stevenson} D.~J.,  2010, \mn@doi [\apjl]
  {10.1088/2041-8205/714/2/L238}, \href
  {http://adsabs.harvard.edu/abs/2010ApJ...714L.238B} {714, L238}

\bibitem[\protect\citeauthoryear{{Charbonneau} \& {Deming}}{{Charbonneau} \&
  {Deming}}{2007}]{2007arXiv0706.1047C}
{Charbonneau} D.,  {Deming} D.,  2007, preprint, \href
  {http://adsabs.harvard.edu/abs/2007arXiv0706.1047C} {} (\mn@eprint {arXiv}
  {0706.1047})

\bibitem[\protect\citeauthoryear{{Checlair}, {Menou}  \& {Abbot}}{{Checlair}
  et~al.}{2017}]{2017ApJ...845..132C}
{Checlair} J.,  {Menou} K.,   {Abbot} D.~S.,  2017, \mn@doi [\apj]
  {10.3847/1538-4357/aa80e1}, \href
  {https://ui.adsabs.harvard.edu/abs/2017ApJ...845..132C} {845, 132}

\bibitem[\protect\citeauthoryear{{Dobbs-Dixon} \& {Agol}}{{Dobbs-Dixon} \&
  {Agol}}{2013}]{2013MNRAS.435.3159D}
{Dobbs-Dixon} I.,  {Agol} E.,  2013, \mn@doi [\mnras] {10.1093/mnras/stt1509},
  \href {http://adsabs.harvard.edu/abs/2013MNRAS.435.3159D} {435, 3159}

\bibitem[\protect\citeauthoryear{Fraedrich, Jansen, Kirk, Luksch  \&
  Lunkeit}{Fraedrich et~al.}{2005}]{PlaSim}
Fraedrich K.,  Jansen H.,  Kirk E.,  Luksch U.,   Lunkeit F.,  2005, \mn@doi
  [Meteorologische Zeitschrift] {doi:10.1127/0941-2948/2005/0043}, 14, 299

\bibitem[\protect\citeauthoryear{{Fromang}, {Leconte}  \& {Heng}}{{Fromang}
  et~al.}{2016}]{2016A&A...591A.144F}
{Fromang} S.,  {Leconte} J.,   {Heng} K.,  2016, \mn@doi [\aap]
  {10.1051/0004-6361/201527600}, \href
  {http://adsabs.harvard.edu/abs/2016A%26A...591A.144F} {591, A144}

\bibitem[\protect\citeauthoryear{{Hammond} \& {Pierrehumbert}}{{Hammond} \&
  {Pierrehumbert}}{2018}]{2018ApJ...869...65H}
{Hammond} M.,  {Pierrehumbert} R.~T.,  2018, \mn@doi [\apj]
  {10.3847/1538-4357/aaec03}, \href
  {https://ui.adsabs.harvard.edu/abs/2018ApJ...869...65H} {869, 65}

\bibitem[\protect\citeauthoryear{{Heng}, {Menou}  \& {Phillipps}}{{Heng}
  et~al.}{2011}]{2011MNRAS.413.2380H}
{Heng} K.,  {Menou} K.,   {Phillipps} P.~J.,  2011, \mn@doi [\mnras]
  {10.1111/j.1365-2966.2011.18315.x}, \href
  {https://ui.adsabs.harvard.edu/abs/2011MNRAS.413.2380H} {413, 2380}

\bibitem[\protect\citeauthoryear{{Komacek} \& {Showman}}{{Komacek} \&
  {Showman}}{2019}]{2019arXiv191009523K}
{Komacek} T.~D.,  {Showman} A.~P.,  2019, arXiv e-prints, \href
  {https://ui.adsabs.harvard.edu/abs/2019arXiv191009523K} {p. arXiv:1910.09523}

\bibitem[\protect\citeauthoryear{{Kuo}}{{Kuo}}{1949}]{1949JAtS....6..105K}
{Kuo} H.-L.,  1949, \mn@doi [Journal of Atmospheric Sciences]
  {10.1175/1520-0469(1949)006<0105:DIOTDN>2.0.CO;2}, \href
  {https://ui.adsabs.harvard.edu/abs/1949JAtS....6..105K} {6, 105}

\bibitem[\protect\citeauthoryear{{Li} \& {Goodman}}{{Li} \&
  {Goodman}}{2010}]{2010ApJ...725.1146L}
{Li} J.,  {Goodman} J.,  2010, \mn@doi [\apj] {10.1088/0004-637X/725/1/1146},
  \href {http://adsabs.harvard.edu/abs/2010ApJ...725.1146L} {725, 1146}

\bibitem[\protect\citeauthoryear{{Liu} \& {Showman}}{{Liu} \&
  {Showman}}{2013}]{2013ApJ...770...42L}
{Liu} B.,  {Showman} A.~P.,  2013, \mn@doi [\apj] {10.1088/0004-637X/770/1/42},
  \href {https://ui.adsabs.harvard.edu/abs/2013ApJ...770...42L} {770, 42}

\bibitem[\protect\citeauthoryear{{Madhusudhan}, {Ag{\'u}ndez}, {Moses}  \&
  {Hu}}{{Madhusudhan} et~al.}{2016}]{2016SSRv..205..285M}
{Madhusudhan} N.,  {Ag{\'u}ndez} M.,  {Moses} J.~I.,   {Hu} Y.,  2016, \mn@doi
  [Space Science Reviews] {10.1007/s11214-016-0254-3}, \href
  {http://adsabs.harvard.edu/abs/2016SSRv..205..285M} {205, 285}

\bibitem[\protect\citeauthoryear{{Mayne} et~al.,}{{Mayne}
  et~al.}{2017}]{2017A&A...604A..79M}
{Mayne} N.~J.,  et~al., 2017, \mn@doi [\aap] {10.1051/0004-6361/201730465},
  \href {https://ui.adsabs.harvard.edu/abs/2017A&A...604A..79M} {604, A79}

\bibitem[\protect\citeauthoryear{{Mendon{\c{c}}a}}{{Mendon{\c{c}}a}}{2019}]{2019arXiv191010760M}
{Mendon{\c{c}}a} J.~M.,  2019, arXiv e-prints, \href
  {https://ui.adsabs.harvard.edu/abs/2019arXiv191010760M} {p. arXiv:1910.10760}

\bibitem[\protect\citeauthoryear{{Menou}}{{Menou}}{2012}]{2012ApJ...745..138M}
{Menou} K.,  2012, \mn@doi [\apj] {10.1088/0004-637X/745/2/138}, \href
  {http://adsabs.harvard.edu/abs/2012ApJ...745..138M} {745, 138}

\bibitem[\protect\citeauthoryear{{Menou}}{{Menou}}{2019}]{2019MNRAS.485L..98M}
{Menou} K.,  2019, \mn@doi [\mnras] {10.1093/mnrasl/slz041}, \href
  {https://ui.adsabs.harvard.edu/abs/2019MNRAS.485L..98M} {485, L98}

\bibitem[\protect\citeauthoryear{{Menou} \& {Rauscher}}{{Menou} \&
  {Rauscher}}{2009}]{2009ApJ...700..887M}
{Menou} K.,  {Rauscher} E.,  2009, \mn@doi [\apj]
  {10.1088/0004-637X/700/1/887}, \href
  {https://ui.adsabs.harvard.edu/abs/2009ApJ...700..887M} {700, 887}

\bibitem[\protect\citeauthoryear{{Paradise} \& {Menou}}{{Paradise} \&
  {Menou}}{2017}]{2017ApJ...848...33P}
{Paradise} A.,  {Menou} K.,  2017, \mn@doi [\apj] {10.3847/1538-4357/aa8b1c},
  \href {https://ui.adsabs.harvard.edu/abs/2017ApJ...848...33P} {848, 33}

\bibitem[\protect\citeauthoryear{{Paradise}, {Menou}, {Valencia}  \&
  {Lee}}{{Paradise} et~al.}{2018}]{2018arXiv180300511P}
{Paradise} A.,  {Menou} K.,  {Valencia} D.,   {Lee} C.,  2018, arXiv e-prints,
  \href {https://ui.adsabs.harvard.edu/abs/2018arXiv180300511P} {p.
  arXiv:1803.00511}

\bibitem[\protect\citeauthoryear{{Parmentier} \& {Crossfield}}{{Parmentier} \&
  {Crossfield}}{2018}]{2018haex.bookE.116P}
{Parmentier} V.,  {Crossfield} I. J.~M.,  2018, {Exoplanet Phase Curves:
  Observations and Theory}.
p.~116, \mn@doi{10.1007/978-3-319-55333-7_116}

\bibitem[\protect\citeauthoryear{{Perna}, {Menou}  \& {Rauscher}}{{Perna}
  et~al.}{2010}]{2010ApJ...719.1421P}
{Perna} R.,  {Menou} K.,   {Rauscher} E.,  2010, \mn@doi [\apj]
  {10.1088/0004-637X/719/2/1421}, \href
  {http://adsabs.harvard.edu/abs/2010ApJ...719.1421P} {719, 1421}

\bibitem[\protect\citeauthoryear{{Rauscher} \& {Menou}}{{Rauscher} \&
  {Menou}}{2012}]{2012ApJ...750...96R}
{Rauscher} E.,  {Menou} K.,  2012, \mn@doi [\apj] {10.1088/0004-637X/750/2/96},
  \href {http://adsabs.harvard.edu/abs/2012ApJ...750...96R} {750, 96}

\bibitem[\protect\citeauthoryear{{Seager} \& {Deming}}{{Seager} \&
  {Deming}}{2010}]{2010ARA&A..48..631S}
{Seager} S.,  {Deming} D.,  2010, \mn@doi [Annual Review of Astron. \& Astrop.]
  {10.1146/annurev-astro-081309-130837}, \href
  {http://adsabs.harvard.edu/abs/2010ARA%26A..48..631S} {48, 631}

\bibitem[\protect\citeauthoryear{{Showman} \& {Polvani}}{{Showman} \&
  {Polvani}}{2011}]{2011ApJ...738...71S}
{Showman} A.~P.,  {Polvani} L.~M.,  2011, \mn@doi [\apj]
  {10.1088/0004-637X/738/1/71}, \href
  {https://ui.adsabs.harvard.edu/abs/2011ApJ...738...71S} {738, 71}

\bibitem[\protect\citeauthoryear{{Showman}, {Gierasch}  \& {Lian}}{{Showman}
  et~al.}{2006}]{2006Icar..182..513S}
{Showman} A.~P.,  {Gierasch} P.~J.,   {Lian} Y.,  2006, \mn@doi [\icarus]
  {10.1016/j.icarus.2006.01.019}, \href
  {https://ui.adsabs.harvard.edu/abs/2006Icar..182..513S} {182, 513}

\bibitem[\protect\citeauthoryear{{Thorngren} \& {Fortney}}{{Thorngren} \&
  {Fortney}}{2018}]{2018AJ....155..214T}
{Thorngren} D.~P.,  {Fortney} J.~J.,  2018, \mn@doi [\aj]
  {10.3847/1538-3881/aaba13}, \href
  {http://adsabs.harvard.edu/abs/2018AJ....155..214T} {155, 214}

\bibitem[\protect\citeauthoryear{{Vallis}}{{Vallis}}{2006}]{2006aofd.book.....V}
{Vallis} G.~K.,  2006, {Atmospheric and Oceanic Fluid Dynamics},
  \mn@doi{10.2277/0521849691.
}

\makeatother
\end{thebibliography}

%\begin{thebibliography}{99}
%\end{thebibliography}

%%%%%%%%%%%%%%%%%%%%%%%%%%%%%%%%%%%%%%%%%%%%%%%%%%

%%%%%%%%%%%%%%%%% APPENDICES %%%%%%%%%%%%%%%%%%%%%

\appendix

\section{PlaSim-Gen Model Description}

PlaSim is an intermediate complexity Earth system simulator that is fast, parallelized, modular and extensively documented \citep{PlaSim}. It has been used and extended before to study the climate of Earth-like exoplanets, with varying degrees of deviation from strict Earth conditions \citep[e.g.][]{2017ApJ...848...33P, 2017ApJ...845..132C,2018arXiv180300511P}. Here we describe how we turned PlaSim into PlaSim-Gen, a generic simulator for deep atmospheres, by removing atmosphere-surface interactions and implementing a simple Newtonian relaxation scheme to drive atmospheric motions.

Our first code-level modification is to redistribute model pressure levels on a logarithmic grid, starting at the bottom pressure level {\tt PSURF} (a standard PlaSim parameter). We followed the exact same procedure as \cite{2012ApJ...750...96R} in distributing sigma levels logarithmically over {\tt OOM} orders of magnitude. Note that PlaSim's dynamical core (PUMA) is a modern, parallelized version of the IGM spectral dynamical core used by \cite{2009ApJ...700..887M, 2012ApJ...750...96R}.

Our second code-level modification is to bypass PlaSim's Earth-centric radiation scheme and  directly implement, in the radiation module, temperature tendencies that obey a 'Newtonian' (linear) relaxation to a prescribed temperature profile, $T_{\rm eq}$, on a prescribed radiative relaxation time, $\tau_{\rm rad}$:

\begin{equation}
\frac{\partial T}{ \partial t} = \frac{T_{\rm eq} - T}{\tau_{\rm rad}}.
\end{equation}
We adopt the fits of \cite{2011MNRAS.413.2380H} for the dayside and nightside profiles of $T_{\rm eq}$ and $\tau_{\rm rad}$, which makes our model specifically tailored to the hot Jupiter HD209458b. 

Our third code-level modification is to change the timestep units in the code so that the parameter {\tt MPSTEP} refers to the (integer) number of seconds per timestep (rather than minutes, as originally implemented). This is to accommodate the possibility of short timestep requirements when modeling hot atmospheres.

We have also removed a few Earth-specific prescriptions that were hardcoded in PlaSim (concerning the magnitude of horizontal hyperdissipation at specific resolutions)

Finally, we use the highly modular character of PlaSim to isolate relevant physics parametrizations while turning off all Earth-specifics, and otherwise unnecessary, modules.  Table~A1 lists the PlaSim parameter choices that allow us to model the deep atmosphere of the hot Jupiter HD209458b and to successfully benchmark against the similar implementation described in \cite{2011MNRAS.413.2380H}.

In short, we select 30 vertical levels (spaced logarithmically), keep dry convective adjustment, choose an appropriate reference temperature and select a short enough timestep for numerical stability. We also adjusted the value of the Robert-Asselin time filter and the type of horizontal hyperdissipation implemented to be more in line with what has been used in other spectral dynamical cores when modeling hot Jupiters \citep[e.g.,][]{2011MNRAS.413.2380H,2012ApJ...750...96R}. In terms of deselecting PlaSim modules, we turn off the sea ice and ocean modules, and we remove surface evaporation, surface fluxes and surface stresses. All precipitations are also turned off, so that our model is effectively dry with zero water content. We turned off PlaSim exisiting Newtonian relaxation scheme since it is super-seeded by the radiation scheme directly applied to temperature tendencies (Equation~A1 above).

Note that some of our parameter choices are also chosen so that a suitable radiative transfer scheme would reproduce a tidally-locked hot Jupiter configuration (e.g. with the diurnal cycle on). Those choices are not strictly necessary given our hard-coding of temperature tendencies with a Newtonian relaxation scheme but they illustrate how PlaSim can be extended to account for permanent dayside insolation with a full radiation scheme \citep[as used in, e.g.,][]{2017ApJ...845..132C}.

\begin{table}
	\centering
	\caption{PlaSim-Gen: Core Model Parameters}
	\label{tab:example_table}
	\begin{tabular}{ccc} % four columns, alignment for each
		\hline
		Parameter & PlaSim-Gen Value & Brief Description\\
		\hline
		NLEV & 30 & Number of (logarithmic) pressure levels \\
		PNU & 0.02 & Robert-Asselin time filter\\ 
		NDCA & 1 & Dry convective adjustment (on)\\ 
		NPRL & 0 & Large scale precipitation (off)\\ 
		NPRC & 0 & Convective precipitation (off)\\ 
		NEVAP & 0 & Surface evaporation (off)\\ 
		NPERPETUAL & 1 & Fixed Orbit (on) \\ 
		NDCYCLE & 1 &  Diurnal cycle (on)\\ 
		NOCEAN & 0 & Ocean model (off)\\ 
		NICE & 0 & Sea ice model (off)\\ 
		NFLUX & 0 & Surface fluxes (off)\\ 
		NSTRESS & 0 & Surface stresses (off)\\ 
		NHDIFF & 0 & Cut off for horizontal diffusion (off)\\ 
		MPSTEP & 180 & Seconds per timestep\\ 
		T0 & $1400$ & Reference temperature (K)\\
		TAUNC & 0 & Timescale for Newtonian cooling (off)\\ 
		& & \\ 
		& & \\ 
		& & \\ 
		\hline
	\end{tabular}
    \label{table:two}
\end{table}

%%%%%%%%%%%%%%%%%%%%%%%%%%%%%%%%%%%%%%%%%%%%%%%%%%

% Don't change these lines
\bsp	% typesetting comment
\label{lastpage}
\end{document}